\documentclass{article}
\usepackage[preprint]{spconf}
\usepackage{graphicx}
\usepackage{mathtools}
\usepackage{amsfonts,amsmath}
\usepackage{booktabs}
\usepackage{url}
\usepackage{pdfpages}

\def\R{{\mathbb R}}

\title{A Deep Representation for Invariance And Music Classification}

\name{Chiyuan Zhang$^{\star}$, Georgios Evangelopoulos$^{\star \dagger}$, Stephen Voinea$^{\star}$, Lorenzo Rosasco$^{\star \dagger}$, Tomaso Poggio$^{\star \dagger}$\thanks{This material is based upon work supported by the Center for Brains, Minds and Machines (CBMM), funded by NSF STC award CCF-1231216. Lorenzo Rosasco acknowledges the financial support of the Italian Ministry of Education, University and Research FIRB project RBFR12M3AC.}}
\address{ $^{\star}$~Center for Brains, Minds and Machines $|$ McGovern Institute for Brain Research at MIT \\
$^{\dagger}$~LCSL, Poggio Lab, Istituto Italiano di Tecnologia and Massachusetts Institute of Technology}

\begin{document}

\includepdf[pages={1}]{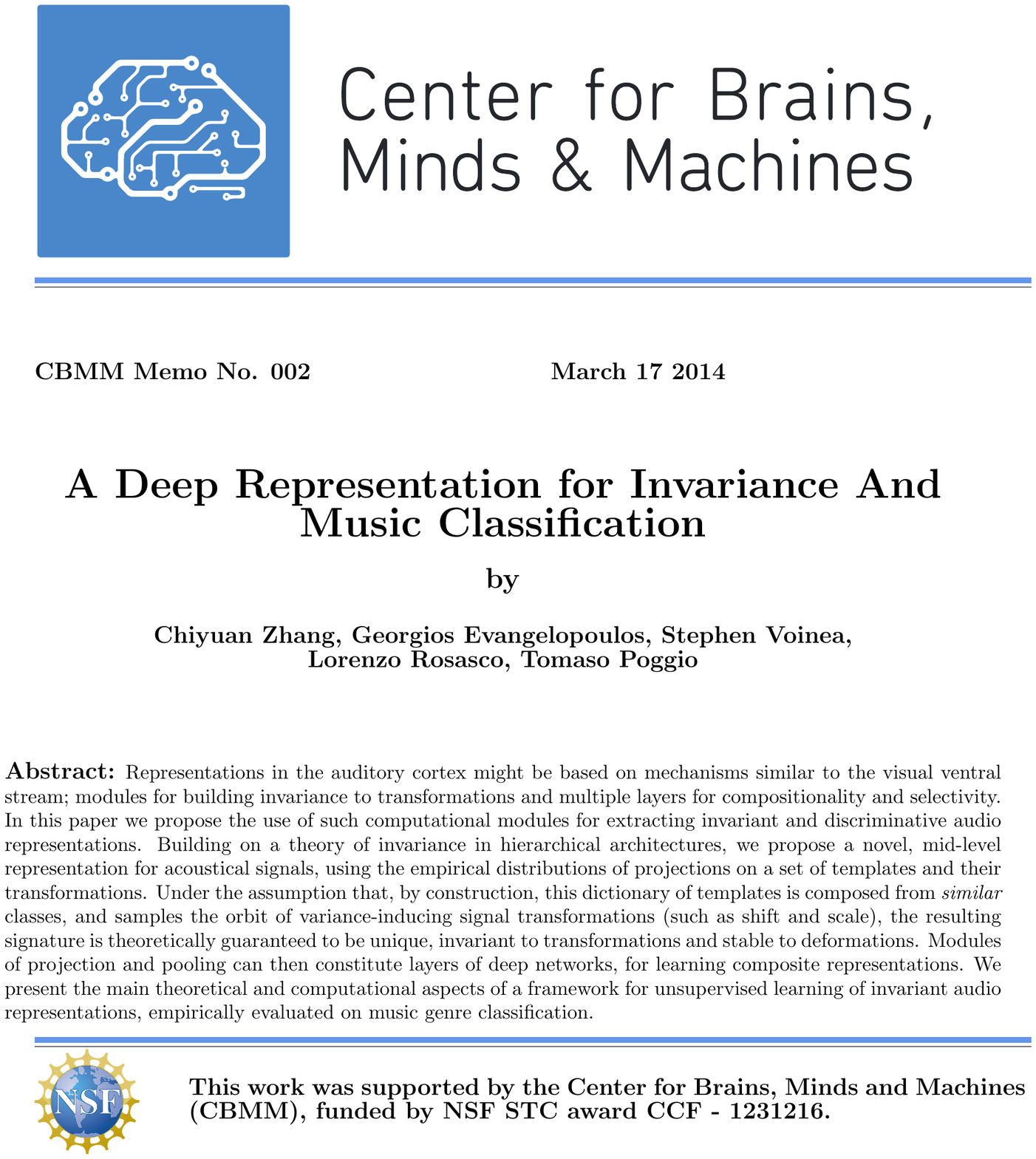}
\setcounter{page}{1}
\toappear{To appear in {\it IEEE 2014 Int'l Conf. on Acoustics, Speech, and Signal Proc. (ICASSP), May 4-9, 2014, Florence, Italy.}}

\maketitle

\begin{abstract}
Representations in the auditory cortex might be based on mechanisms similar to the visual ventral stream; modules for building invariance to transformations and 
multiple layers for compositionality and selectivity. In this paper we propose the use of such computational modules for extracting invariant and discriminative audio representations. Building on a theory of invariance in hierarchical architectures, we propose a novel, mid-level representation for acoustical signals, using the empirical distributions of projections on a set of templates and their transformations. Under the assumption that, by construction, this dictionary of templates is composed from \emph{similar} classes, and samples the orbit of variance-inducing signal transformations (such as shift and scale), the resulting signature is theoretically guaranteed to be unique, invariant to transformations and stable to deformations. Modules of projection and pooling can then constitute layers of deep networks, for learning composite representations. We present the main theoretical and computational aspects of a framework for unsupervised learning of invariant audio representations, empirically evaluated on music genre classification.
\end{abstract}

\begin{keywords}
Invariance, Deep Learning, Convolutional Networks, Auditory Cortex, Music Classification
\end{keywords}
%
\section{Introduction}
\label{sec:intro}

The representation of music signals, with the goal of learning for recognition, classification, context-based recommendation, annotation and tagging, mood/theme detection, summarization etc., has been relying on techniques from speech analysis. For example, Mel-Frequency Cepstral Coefficients (MFCCs), a widely used representation in automatic speech recognition, is computed from the Discrete Cosine Transform of Mel-Frequency Spectral Coefficients (MFSCs). The assumption of signal stationarity within an analysis window is implicitly made, thus dictating small signal segments (typically 20-30ms) in order to minimize the loss of non-stationary structures for phoneme or word recognition. 
Music signals involve larger scale structures though (on the order of seconds) that encompass discriminating features, apart from musical timbre, such as melody, harmony, phrasing, beat, rhythm etc.

The acoustic and structural characteristics of music have been shown to require a distinct characterization of structure and content \cite{Muller2011}, and quite often a specialized feature design.
A recent critical review of features for music processing \cite{Humphrey2013} identified three main shortcomings: a)~the lack of scalability and generality of task-specific features, b)~the need for higher-order functions as approximations of nonlinearities, c)~the discrepancy between short-time analysis with larger, temporal scales where music content, events and variance reside. 

Leveraging on a theory for invariant representations \cite{magic} and an associated computational model of hierarchies of projections and pooling, 
we propose a hierarchical architecture that learns a representation invariant to transformations and stable \cite{CPA:CPA21413}, over large analysis frames. 
We demonstrate how a deep representation, invariant to typical transformations, improves music classification and how unsupervised learning is feasible using stored templates and their transformations.


\section{Related Work}

Deep learning and convolutional networks (CNNs) have been recently applied for learning mid- and high- level audio representations, motivated by successes in improving image and speech recognition. Unsupervised, hierarchical audio representations from Convolutional Deep Belief Networks (CDBNs) have improved music genre classification over MFCC and spectrogram-based features \cite{Lee:2009wl}. Similarly, Deep Belief Networks (DBNs) were applied for learning music representations in the spectral domain \cite{Hamel2010} and unsupervised, sparse-coding based learning for audio features \cite{Henaff2011}.


A mathematical framework that formalizes the computation of invariant and stable representations via cascaded (deep) wavelet transforms has been proposed in \cite{CPA:CPA21413}. In this work, we propose computing an audio representation through biologically plausible modules of projection and pooling, based on a theory of invariance in the ventral stream of the visual cortex \cite{magic}. 
The proposed representation can be extended to hierarchical architectures of ``layers of invariance". An additional advantage is that it can be applied to building invariant representations from arbitrary signals without explicitly modeling the underlying transformations, which can be arbitrarily complex but smooth.



Representations of music directly from the temporal or spectral domain can be very sensitive to small time and frequency deformations, which affect the signal but not its musical characteristics. In order to get stable representations, pooling (or aggregation) over time/frequency is applied to smooth-out such variability. Conventional MFSCs use filters with wider bands in higher frequencies to compensate for the instability to deformations of the high-spectral signal components. The scattering transform \cite{DBLP:conf/ismir/AndenM11,DBLP:journals/corr/abs-1304-6763} keeps the low pass component of cascades of wavelet transforms as a layer-by-layer average over time. 
Pooling over time or frequency is also crucial for CNNs applied to speech and audio  \cite{Lee:2009wl, AbdelHamid:2012hs}.

\section{Unsupervised Learning of Invariant Representations}
\label{sec:magic-theory}
Hierarchies of appropriately tuned neurons can compute stable and invariant representations using only primitive computational operations of high-dimensional inner-product and nonlinearities \cite{magic}. We explore the computational principles of this theory in the case of audio signals and propose a multilayer network for invariant features over large windows. 

\subsection{Group Invariant Representation}

Many signal transformations, such as shifting and scaling can be naturally modeled by the action of a group $G$. We consider 
transformations that form a compact group, though, as will be shown, the general theory holds (approximately) for a much more general class (e.g., smooth deformations). Consider a segment of an audio signal $x \in \R^d$. For a representation
$\mu(x)$ to be invariant to transformation group $G$, $\mu(x) = \mu(g x)$ has to hold $\forall
g\in G$. The \emph{orbit} $O_x$ is the set of transformed
signals $g x, \forall
g\in G$ generated from the action of the group on $x$, i.e., $O_x=\{gx\in\R^d, g\in G \}$. Two signals $x$ and $x'$ are \emph{equivalent} if they are in the same orbit, that is, $\exists g\in G$, such that $gx = x'$. This equivalence
relation formalizes the \emph{invariance} of the orbit. On the other hand, the orbit
is \emph{discriminative} in the sense that if $x'$ is not a transformed
version of $x$, then orbits $O_x$ and $O_{x'}$ should be different.

Orbits, although a convenient mathematical formalism, are difficult to work with
in practice. When $G$ is compact, we can normalize the Haar measure on
$G$ to get an induced probability distribution $P_x$ on the transformed signals, 
which is also invariant and discriminative.
The high-dimensional distribution $P_x$ 
can be estimated within small $\epsilon$ in terms of the set of one dimensional distributions induced from projecting $gx$ onto vectors on the unit sphere, following Cram{\'e}r-Wold Theorem \cite{cramer-wold} and concentration of measures \cite{magic}.  
Given a finite set of randomly-chosen, unit-norm templates $t^1,\ldots,t^K$, an invariant signature for $x$ is approximated by the set of $P_{\langle x, t^k\rangle}$, by computing $\langle gx, t^k\rangle, \forall g\in G, k=1,\ldots,K$ and estimating the one dimensional histograms $\mu^k(x) = (\mu^k_n(x))_{n=1}^N$. For a (locally) compact group $G$,
\begin{equation}
	\mu_n^k(x) = \int_G \eta_n\left(\langle gx, t^k\rangle \right) dg
	\label{eq:complex-cell}
\end{equation}
is the $n$-th histogram bin of the distribution of projections onto the $k$-th template, implemented by the nonlinearity $\eta_n(\cdot)$.
The final representation $\mu(x)\in \R^{NK}$ is the concatenation of the $K$ histograms.

Such a signature is impractical because it requires access to all transformed versions $gx$ of the input $x$. The simple property $\langle gx, t^k\rangle = \langle x, g^{-1}t^k\rangle$, allows for a memory-based learning of invariances; instead of all transformed versions of input $x$, the neurons can store all transformed versions of all the templates $gt^k$, $g\in G, k=1,\ldots, K$ during training. 
The implicit knowledge stored in the transformed templates allows for the computation of invariant signatures without \emph{explicit understanding} of the underlying transformation group. 

For the visual cortex, the templates and their transformed versions could be learned from unsupervised visual experience through Hebbian plasticity \cite{HebbianPlasticity}, assuming temporally adjacent images would typically correspond to (transformations of) the same object. Such memory-based learning might also apply to the auditory cortex and audio templates could be observed and stored in a similar way. In this paper, we sample templates randomly from a training set and transform them explicitly according to known transformations.

\subsection{Invariant Feature Extraction with Cortical Neurons}

\begin{figure}[t]
\centering
\includegraphics[width=\linewidth]{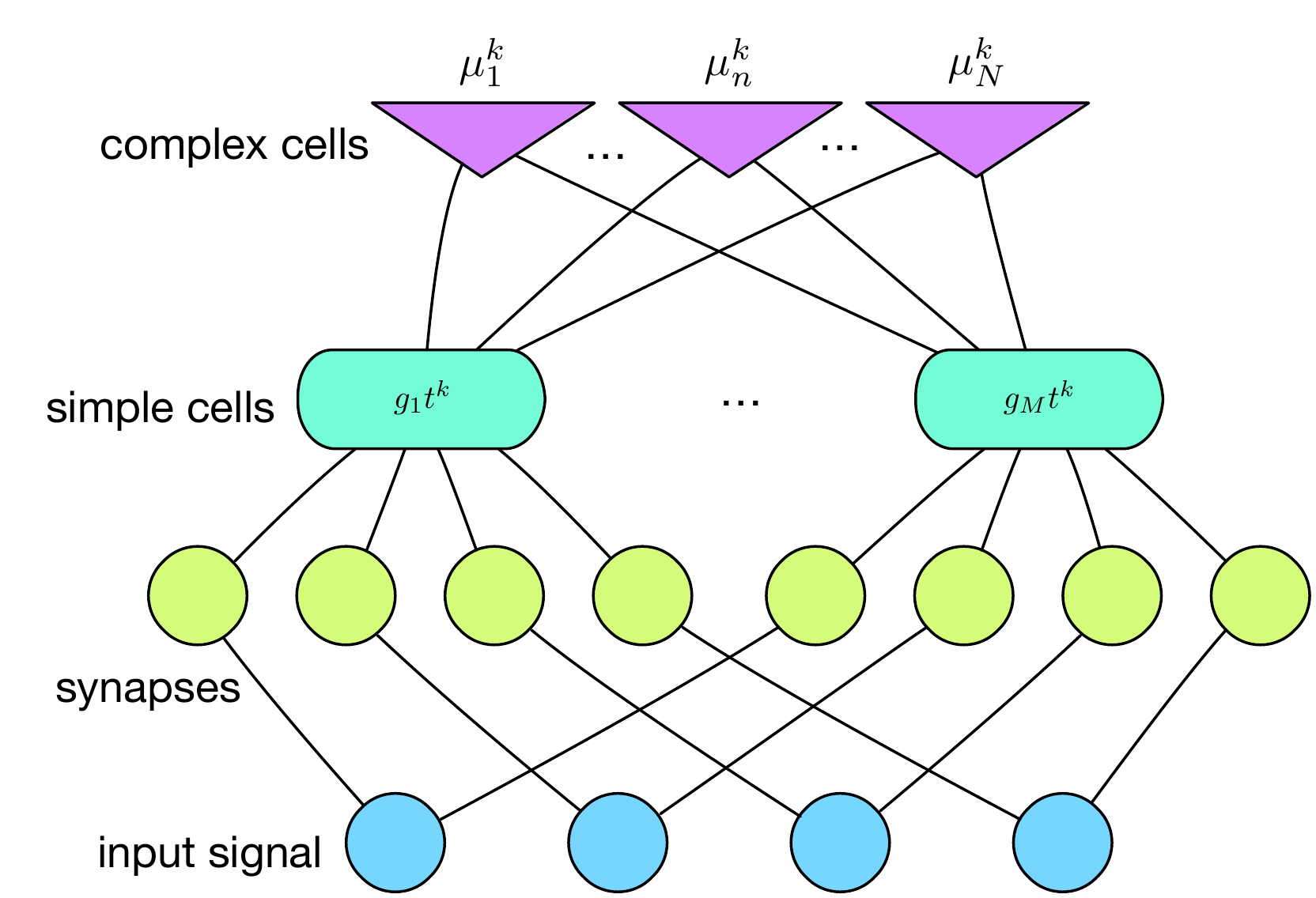}
\vspace{-0.5cm}
\caption{Illustration of a simple-complex cell module (projections-pooling) that computes an invariant signature component for the $k$-th template.}
\label{fig:simple-complex}
\end{figure}

The computations for an invariant representation can be carried out by primitive neural operations. The cortical neurons typically have $10^3\sim 10^4$ synapses, in which the templates can be stored in the form of synaptic weights. By accumulating signals from synapses, a single neuron can compute a high-dimensional dot-product between the input signal and a transformed template.

Consider a module of \emph{simple} and \emph{complex} cells \cite{HubelWiesel} associated with template $t^k$, illustrated in Fig.~\ref{fig:simple-complex}. Each simple cell stores in its synapses a single transformed template $g_1t^k,\ldots,g_Mt^k$, where $M = |G|$. For an input signal, the cells compute the set of inner products $\{\langle x, gt^k\rangle\}, \forall g\in G$. Complex cells accumulate those inner products and pool over them using a nonlinear function $\eta_n(\cdot)$. 
For families of smooth step functions (sigmoids) 
\begin{equation}
	\eta_n(\cdot) = \sigma(\cdot + n\Delta),
\end{equation}
the $n$-th cell could compute the $n$-th bin of an empirical Cumulative Distribution Function for the underlying distribution Eq.~\eqref{eq:complex-cell}, with $\Delta$ controlling the size of the histogram bins.

Alternatively, the complex cells could compute moments of the distribution, with $\eta_n(\cdot) = (\cdot)^n$ corresponding to the $n$-th order moment. Under mild assumptions, the moments could be used to approximately characterize the underlying distribution. Since the goal is an invariant signature instead of a complete distribution characterization, a finite number of moments would
suffice. Notable special cases include the \emph{energy model} of complex cells \cite{energyComplexCell} for $n=2$ and \emph{mean pooling} for $n=1$. 

The computational complexity and approximation accuracy (i.e., finite samples to approximate smooth transformation groups and discrete histograms to approximate continuous distributions) grows linearly with the number of transformations per group and number of histogram bins. In the computational model these correspond to number of simple and complex cells, respectively, and can be carried out in parallel in a biological or any parallel-computing system.   


\subsection{Extensions: Partially Observable Groups and Non-group Transformations}

For groups that are only observable within a \emph{window} over the orbit, i.e. partially observable groups, or pooling over a subset of a finite group $G_0\subset G$ (not necessarily a \emph{subgroup}), a local signature associated with $G_0$ can be computed as
\begin{equation}
	\mu_n^k(x) = \frac{1}{V_0}\int_{G_0} \eta_n \left( \langle gx, t^k\rangle  \right) dg
\end{equation}
where $V_0 = \int_{G_0} dg$ is a normalization constant to define a valid
probability distribution. It can be shown that this representation is
\emph{partially invariant} to a restricted subset of transformations \cite{magic},  
if the input and templates have a \emph{localization property}. The case for general (non-group) smooth transformations is more complicated. The smoothness assumption implies that local linear approximations centered around some \emph{key transformation parameters} are possible, and for local neighborhoods, the POG signature properties imply approximate invariance \cite{magic}.


\section{Music Representation and Genre Classification}

\begin{figure*}[tb]
\centering
\includegraphics[width=0.9\textwidth]{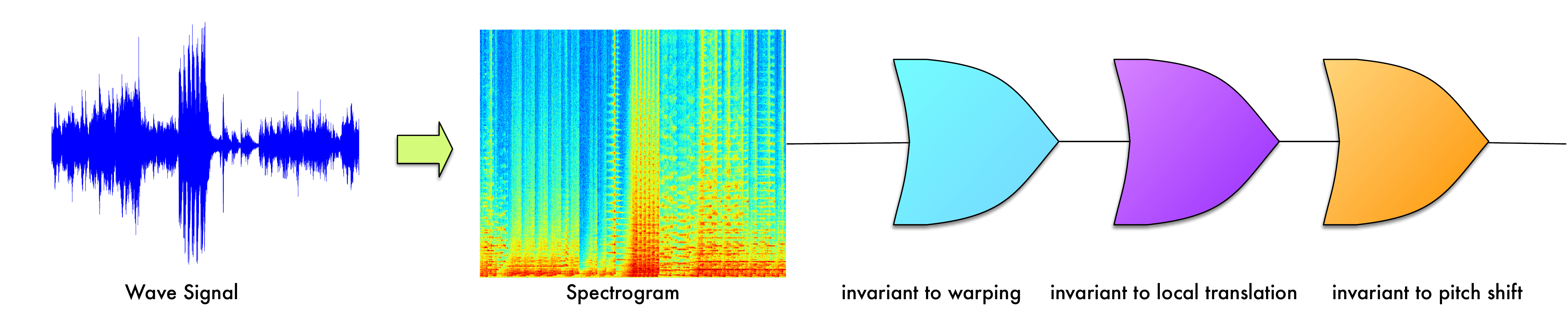}
\vspace{-0.4cm}
\caption{Deep architecture for invariant feature extraction with cascaded transform invariance layers.}
\label{fig:pipeline}
\end{figure*}

The repetition of the main module on multilayer, recursive architectures, can 
build layer-wise invariance of increasing range and an \emph {approximate factorization} of stacked transformations. In this paper, we focus on the latter and propose a multilayer architecture for a deep representation and feature extraction, illustrated in Fig.~\ref{fig:pipeline}. Different layers are tuned to impose invariance to audio changes such as warping, local translations and pitch shifts. We evaluate the properties of the resulting audio signature for musical genre classification, 
by cascading layers while comparing to ``shallow'' (MFCC) and ``deep'' (Scattering) representations.

\vspace{-0.2cm}
\subsection{Genre Classification Task and Baselines}

The GTZAN dataset \cite{DBLP:journals/taslp/TzanetakisC02} consists of 1000
audio tracks each of 30 sec length, some containing vocals, that are evenly divided into 10 music genres. 
To classify tracks into genres using frame level features, we follow a frame-based, majority-voting scheme \cite{DBLP:conf/ismir/AndenM11}; each frame is classified independently and a global label is assigned using majority voting over all track frames. 
To focus on the discriminative strength of the representations, 
we use one-vs-rest multiclass reduction with regularized linear least squares as base classifiers \cite{GURLS}. 
The dataset is randomly split into a 80:20 partition of train and test data. 


Results for genre classification are shown in Table~\ref{tab:gtzan}. As a baseline, MFCCs computed over longer (370 ms) windows achieve a track error rate of 67.0\%. Smaller-scale MFCCs 
can not capture long-range structures and under-perform when applied to music genre classification \cite{DBLP:conf/ismir/AndenM11}, while longer windows violate the assumption of signal stationarity, leading to large information loss. The scattering transform adds layers of wavelet convolutions and modulus operators to recover the non-stationary behavior lost by MFCCs \cite{CPA:CPA21413, DBLP:conf/ismir/AndenM11,DBLP:journals/corr/abs-1304-6763}; it is both translation-invariant and stable to time warping. A second-order scattering transform representation, 
greatly decreases the MFCC error rate at 24.0\% 
The addition of higher-order layers improves the performance, but only slightly. 


State-of-the-art results for the genre task combine multiple features and well-adapted classifiers. On GTZAN\footnote{Since there is no standard training-testing partition of the GTZAN dataset, error rates may not be directly comparable.}, a 9.4\% error rate is obtained by combining MFCCs with stabilized modulation spectra \cite{Chang2009}; combination of cascade filterbanks with sparse coding yields a 7.6\% error \cite{Panagakis09musicgenre}; scattering transform achieves an error of 8.1\% when combining adaptive wavelet octave bandwidth, multiscale representation and all-pair nonlinear SVMs \cite{DBLP:journals/corr/abs-1304-6763}. 

\vspace{-0.3cm}
\subsection{Multilayer Invariant Representation for Music}
\label{sec:invariant-gtzan}


At the {\bf base layer}, we compute a log-spectrogram representation using a short-time Fourier transform in 370 ms windows, in order to capture long-range audio signal structure. As shown Table~\ref{tab:gtzan}, the error rate from this input layer alone is 35.5\%, better than MFCC, but worse than the scattering transform. This can be attributed to the instability of the spectrum to time warping at high frequencies \cite{DBLP:journals/corr/abs-1304-6763}. 

Instead of average-pooling over frequency, as in a mel-frequency transformation (i.e., MFSCs), we handle instability using mid-level representations built for invariance to warping (Sec.~\ref{sec:magic-theory}). Specifically, we add a {\bf second layer} to pool over projections on warped templates on top of the spectrogram layer. The templates are audio segments randomly sampled from the training data. For each template $t^k[n]$, we explicitly warp the signal as $g_\epsilon t^k[n] = t^k_\epsilon [n] = t^k[(1+\epsilon)n]$ for a large number of $\epsilon \in [-0.4, 0.4]$. We compute the normalized dot products between input and templates (projection step), collect values for each template group $k$ and estimate the first three moments of the distribution for $k$ (pooling step). The representation $(\mu^k(x))_1^K$ at this layer is then the concatenation of moments from all template groups. An error rate of 22.0\% is obtained with this representation, a significant improvement over the base layer representation, that notably outperforms both the 2nd and 3rd order scattering transform.

In a {\bf third layer}, we handle local translation invariance by explicitly \emph{max pooling} over neighboring frames. A neighborhood of eight frames is pooled via a component-wise max operator. To reduce the computational complexity, we do subsampling by shifting the pooling window by three frames. This operation, similar to the spatial pooling in HMAX \cite{HMAX} and CNNs \cite{Lee:2009wl,AbdelHamid:2012hs,Hamel2011}, could be seen as a special case in our framework: a receptive field covers neighboring frames with max pooling; each template corresponds to an impulse in one of its feature dimensions and templates are translated in time. With this third layer representation, the error rate is further reduced to 16.5\%.

A {\bf fourth layer} performs projection and pooling over pitch-shifted templates, in their third-layer representations, randomly sampled from the training set. Although the performance drops slightly to 18.0\%, it is still better than the compared methods. This drop may be related to several open questions around hierarchical architectures for invariance: 
a)~should the classes of transformations be adapted to specific domains, e.g., the invariant to pitch-shift layer, while natural for speech signals, might not be that relevant for music signals; b)~how can one learn the transformations or obtain the transformed templates automatically from data (in a supervised or unsupervised manner); c)~how many layers are enough when building hierarchies; d)~under which conditions can different layers of invariant modules be stacked.  

The theory applies nicely in a one-layer setting. Also when the transformation (and signature) of the base layer is \emph{covariant} to the upper layer transformations, a hierarchy could be built with provable invariance and stability \cite{magic}. However, \emph{covariance} is usually a very strong assumption in practice. Empirical observations such as these can provide insights on weaker conditions for deep representations with theoretical guarantees on invariance and stability.

%

\begin{table}\centering
\begin{tabular}{lc}
\toprule
Feature & Error Rates (\%) \\
\midrule
MFCC & 67.0 \\
\midrule
Scattering Transform (2nd order) & 24.0 \\
Scattering Transform (3rd order) & 22.5 \\
Scattering Transform (4th order) & 21.5 \\
\midrule
Log Spectrogram & 35.5 \\
Invariant (Warp) & \textbf{22.0} \\
Invariant (Warp+Translation) & \textbf{16.5} \\
Invariant (Warp+Translation+Pitch) & \textbf{18.0} \\
\bottomrule
\end{tabular}
\vspace{-0.22cm}
\caption{Genre classification results on GTZAN with one-vs-rest reduction and \emph{linear} ridge regression binary classifier.}
\label{tab:gtzan}
\end{table}

\vspace{-0.3cm}
\section{Conclusion}

The theory of stacking invariant modules for a hierarchical, deep network is still under active development.  
Currently, rather strong assumptions are needed to guarantee an invariant and stable representation when multiple layers are stacked, and open questions involve the type, number, observation and storage of the transformed template sets (learning, updating etc.). Moreover, systematic evaluations remain to be done for music signals and audio representations in general. Towards this, we will test the performance limits of this hierarchical framework on speech and other audio signals and validate the representation capacity and invariance properties for different recognition tasks. Our end-goal is to push the theory towards a concise prediction of the role of the auditory pathway for unsupervised learning of invariant representations and a formally optimal model for deep, invariant feature learning.



\vfill\pagebreak

\bibliographystyle{IEEEbib}
\bibliography{CBMM_Memo_002}

\end{document}